# Response: "Commentary: Is the moon there if nobody looks? Bell inequalities and physical reality"

Marian Kupczynski*

Département d'informatique et d'ingénierie, Université du Québec en Outaouais (UQO), Gatineau, QC, Canada

A Commentary on
"Commentary: Is the moon there if nobody looks: Bell inequalities and physical reality"

by Gill R. D. and Lambare J. P. (2023). Front. Phys. 10:1024718. doi: 10.3389/fphy.2022.1024718

## 1 Introduction

In [1, 2], Gill and Lambare (GL) criticize our uncontroversial paper [3] published in Frontiers in Physics. They claim that their "*findings*" apply to our preceding papers [4–6] and even to those by other authors. They seem to suggest that the content and conclusions of [3] are invalid because they are based on a false mathematical claim and reasoning. This is unfounded, unfair, and misleading. A longer reply to their criticism may be found in [7].

In [3], we correctly demonstrate that the violation of inequalities and apparent violation of no-signaling in Bell tests may be explained in a locally causal way. Therefore, the violation of inequalities does not allow for doubt regarding the existence of objective external physical reality and causal locality in nature.

The joint probability (JP) distribution of four random variables describes a random experiment in which four outcomes are outputted in each trial. Only then can CHSH be derived and obeyed by all finite samples.

In our probabilistic framework, such JP does not exist, and four experiments are described by random variables implemented on four disjoint dedicated probability spaces.

GL construct a counterfactual probabilistic model in which random variables representing outcomes of four experiments performed using incompatible experimental settings are jointly distributed. Thus, CHSH inequalities trivially hold for all finite samples generated by their model. Their model defines a probabilistic coupling for our model describing raw data from Bell tests [7]. The existence of this coupling does not invalidate the derivation of the contextual probabilistic model describing the final data from Bell tests. Only these final data are used to test Bell inequalities. Inequalities cannot be derived because our model violates *statistical independence* [4,8,10].

## 2 Locally causal description of Bell tests

Statistical inference is based on finite experimental samples. Inequalities can be violated by pseudo-random samples generated using various probabilistic models, including local realistic models (see an excellent review by Larsson[11]). They are violated by experimental data in physics and cognitive science. The important questions we wanted to answer in [3] are

1) Can we explain the data from Bell tests without evoking quantum non-locality and quantum magic?
2) What metaphysical conclusions, if any, may be made if CHSH inequalities are violated in a given experiment?

Raw data from Bell tests are obtained by converting two distant time-series of clicks into samples containing paired outcomes (a, b), with a = ±1 or 0 and b = = ±1 or 0, coding clicks in some synchronized time windows. From raw data, final data are extracted with only non-vanishing pairs (a, b), and pairwise expectations of random variables may be described as conditional expectations [5, 9, 10]:

$$E(A_x B_y | A_x \neq 0, B_y \neq 0) = \sum_{\lambda \in \Lambda'_{xy}} A_x(\lambda_1, \lambda_x) B_y(\lambda_2, \lambda_y) p_x(\lambda_x) p_y(\lambda_y) p(\lambda_1, \lambda_2),$$
(1)

where $\Lambda_{xy} = \Lambda_{12} \times \Lambda_x \times \Lambda_y$ and $\Lambda'_{xy} = \{\lambda \in \Lambda_{xy} | A_x(\lambda_1, \lambda_x) \neq 0 \text{ and } B_y(\lambda_2, \lambda_y) \neq 0\}$. It explains, in a locally causal way, the apparent violation of no-signaling reported in [12–16]:

$$E(A_x | A_x B_y \neq 0) \neq E(A_x | A_x B_{y'} \neq 0); E(B_y | A_x B_y \neq 0) \neq E(B_y | A_{x'} B_y \neq 0).$$
(2)

A procedure for extracting non-vanishing paired outcomes is not unambiguous and is setting-dependent. Therefore, discussing the detection loophole is misleading. One should rather discuss the *photon identification loophole* [17, 18].

Because of an apparent violation of no-signaling (2), in the contextuality-by-default (CbD) approach of Dzhafarov and Kujala [19–21], final data from Bell tests are described by eight binary random variables ($A_{xy}$, $B_{xy}$, $A_{x'y}$, $B_{x'y}$, $A_{xy'}$, $B_{xy'}$, $A_{x'y'}$, $B_{x'y'}$), instead of four variables, and pairwise expectations are evaluated using a new probabilistic model [9]:

$$E(A_{xy} B_{xy}) = \sum_{\lambda \in \Lambda_{xy}} A_x(\lambda_1, \lambda_x) B_y(\lambda_2, \lambda_y) p_{xy}(\lambda_x, \lambda_y) p(\lambda_1, \lambda_2),$$
(3)

where $A_{xy} = \pm 1$ and $B_{xy} = \pm 1$. It is clear that neither the GL probabilistic model nor Bell averaging over instrument variables may be used to prove CHSH inequalities for random experiments described by the probabilistic models (1,3).

Correlations between distant outcomes in Bell tests, often called *non-local*, may be explained using models (1,3). The experimental protocol used in (1, 2) is consistent with the experimental protocol of Weihs et al. [22].

The Delft experiment [23] used a different experimental protocol, but the use of time windows and post-selection could not be avoided [5, 15, 16]. As we explained in [4], "*entanglement swapping*" may also be understood without evoking quantum magic. Contrary to what Aspect claimed, namely that "*Mixing two photons on a beam splitter and detecting them in coincidence entangles the electron spins on the remote NV centers*" [24], the observation of a particular coincidence signal gives only the information that "correlated signals" in distant laboratories were created and measurements were carried out in specific synchronized time slots [4].

## 3 Conclusion

There are no false mathematical claims and false assertions in our paper [3], *around which our work is built*. Signals arriving at measuring stations are described by setting independent random variables, which are statistically dependent and causally independent. Measuring instruments are described by random variables, which are setting-dependent [8, 10]. They are causally independent, but they may be statistically dependent (1,3). We are not looking for an escape route for local realism.

Hidden variables describing measuring instruments are explicitly incorporated in the models (1,3). Thus, they do not suffer from a theoretical *contextual loophole* [25, 26]. Setting dependence of a hidden variable has nothing to do with the lack of free will and should be called *contextuality* [8–10].

Metaphysical conclusions which may be drawn from the violation of inequalities in Bell tests are quite limited [3, 27]. The violation of inequalities does not prove the completeness of QM, which was the subject of the Bohr–Einstein quantum debates [4]. A contextual character of quantum observables and the active role played by measuring instruments were explained by Bohr many years ago. Speculations about *quantum non-locality* are rooted in incorrect interpretations of QM and/or in incorrect "mental pictures" and models trying to provide a more detailed explanation of quantum phenomena [3, 28–32].

The violation of inequalities and apparent violation of non-signaling in Bell tests may be explained in a locally causal way without evoking quantum magic.

Nevertheless, the research stimulated by Bell–CHSH inequalities [33] and the beautiful experiments designed and performed to test them, rewarded recently with a Nobel Prize, paved the way for important applications of "non-local quantum correlations" in quantum information and quantum technologies.

## Author contributions

The author confirms being the sole contributor of this work and has approved it for publication.

## Conflict of interest

The author declares that the research was conducted in the absence of any commercial or financial relationships that could be construed as a potential conflict of interest.